\documentclass{sig-alternate-05-2015}

\usepackage{amsmath}
\usepackage{graphicx, subfig}
\usepackage{caption}
\usepackage{booktabs}
\usepackage{placeins}
\usepackage{multirow}
\usepackage{url}
\usepackage{microtype}
\usepackage{algorithmicx,algorithm,algpseudocode}
\usepackage[export]{adjustbox}
\usepackage{enumitem}
\usepackage{pdfpages}
\usepackage{verbatim}
\usepackage{cite}

\usepackage{ifthen,color}
\definecolor{darkgreen}{rgb} {0.0,0.5,0.0}
\definecolor{darkblue}{rgb} {0.0,0.0,0.5}
\definecolor{bluegreen}{rgb}{0.0,0.5,0.5}
\definecolor{redgreen}{rgb}{0.5,0.5,0.0}

\newcommand{\yzh}[1]{{\it{\color{darkblue}(YZH) #1}}}
\newcommand{\john}   [1]{{{\color{darkgreen}(John) #1}}}

\newcommand{\final}{0}
\ifthenelse{\equal{\final}{1}}{\renewcommand{\john}[1]{}\renewcommand{\yzh}[1]{}}{}
\newcommand{\includehelptoggle}{0}
\ifthenelse{\equal{\includehelptoggle}{1}}{\newcommand{\includehelp}[1]{#1}}{\newcommand{\includehelp}[1]{}}
\ifdefined\textln\relax\else\newcommand{\textln}[1]{#1}\fi

\usepackage{listings}
\usepackage{color}

\definecolor{dkgreen}{rgb}{0,0.6,0}
\definecolor{gray}{rgb}{0.5,0.5,0.5}
\definecolor{mauve}{rgb}{0.58,0,0.82}

\begin{document}

% Copyright
%\setcopyright{acmcopyright}
\setcopyright{acmlicensed}
%\setcopyright{rightsretained}
%\setcopyright{usgov}
%\setcopyright{usgovmixed}
%\setcopyright{cagov}
%\setcopyright{cagovmixed}

% DOI
\doi{http://dx.doi.org/10.1145/2915516.2915521}

% ISBN
\isbn{978-1-4503-4350-3/16/05}\acmPrice{\$15.00}

%Conference
\conferenceinfo{HPGP'16,}{May 31, 2016, Kyoto, Japan.}

%
% --- Author Metadata here ---
%\conferenceinfo{WOODSTOCK}{'97 El Paso, Texas USA}
\CopyrightYear{2016} % Allows default copyright year (20XX) to be over-ridden - IF NEED BE.
%\crdata{0-12345-67-8/90/01}  % Allows default copyright data (0-89791-88-6/97/05) to be over-ridden - IF NEED BE.
% --- End of Author Metadata ---
\clubpenalty = 10000
\widowpenalty = 10000
%Try the first version of the name. Very boring.
\title{A Comparative Study on Exact Triangle Counting Algorithms on the {GPU}}

\numberofauthors{4} %  in this sample file, there are a *total*
\author{
% 1st. author
\alignauthor
Leyuan Wang\\
       \affaddr{University of California, Davis}\\
       \email{leywang@ucdavis.edu}
% 2nd. author
\alignauthor
Yangzihao Wang\\
       \affaddr{University of California, Davis}\\
       \email{yzhwang@ucdavis.edu}
% 3rd. author
\alignauthor
Carl Yang\\
       \affaddr{University of California, Davis}\\
       \email{ctcyang@ucdavis.edu}
\and
\alignauthor
% 4th. author
John D. Owens\\
       \affaddr{University of California, Davis}\\
       \email{jowens@ece.ucdavis.edu}
}

\maketitle
\begin{abstract}
We implement exact triangle counting in graphs on the GPU using three
different methodologies: subgraph matching to a triangle pattern;
programmable graph analytics, with a set-intersection approach; and a
matrix formulation based on sparse matrix-matrix multiplies. All three
deliver best-of-class performance over CPU implementations and over
comparable GPU implementations, with the graph-analytic approach
achieving the best performance due to its ability to exploit efficient
filtering steps to remove unnecessary work and its high-performance
set-intersection core.
\end{abstract}



%
%  Use this command to print the description
%

%\keywords{Graph Processing; Triangle Counting; GPU Computing; Parallel}

\section{Introduction}
\label{sec:intro}
In recent years, graphs have been used to model interactions between entities in a broad spectrum of applications. Graphs can represent relationships in social media, the World Wide Web, biological and genetic interactions, co-author networks, citations, etc. Understanding the underlying structure of these graphs is becoming increasingly important, and one of the key techniques for understanding is based on finding small subgraph patterns.

The most important such subgraph is the triangle. Many important measures of a graph are triangle-based, such as the clustering coefficient and the transitivity ratio. The clustering coefficient is frequently used in measuring the tendency of nodes to cluster together as well as how much a graph resembles a small-world network~\cite{Watts:1998:CDS}. The transitivity ratio is the probability of wedges (three connected nodes) forming a triangle.

A side product obtained from many triangle counting algorithms is triangle enumeration. Instead of solving the $k$-clique problem, which is NP-complete, enumerating triangles is useful as a subroutine in solving $k$-truss, which is a relaxation of the $k$-clique problem~\cite{Wang:2012:TDM}. The three triangle counting algorithms we examine in this paper will also enumerate triangles, so they can be used to solve $k$-truss. Triangle enumeration is useful for dense neighborhood discovery too~\cite{Wang:2010:OTB}.

To count triangles, numerous methods have been proposed from various domains including algorithms based on matrix multiplication~\cite{Coppersmith:1987:MMV}, a MapReduce implementation~\cite{Kolda:2013:CTM}, and many approximation techniques such as wedge sampling~\cite{Talya:2015:ACT,Madhav:2012:FBP,Tsourakakis:2009:DCT}. An experimental study on different sequential algorithms is proposed by Schank and Wagner~\cite{Schank:2005:FCL}. With the explosion in data sizes and the emergence of commodity data-parallel processors, the research community has begun to develop efficient parallel implementations~\cite{Green:2014:FTC,Polak:2015:CTL,Azad:2015:PTC}. In this paper, we focus on data-parallel exact triangle counting algorithms that are suitable for implementations on devices such as many-core GPUs and multi-core CPUs. These processors provide ample computing power for solving data-intensive problems at the cost of limitations imposed by their programming models.

In this work, we do a comparative study on three specialized parallel triangle counting algorithms with highly scalable implementations on NVIDIA GPUs. We make the following contributions:

\begin{enumerate}
	
        \item We survey triangle counting algorithms that are viable for GPUs and apply a state-of-the-art subgraph matching algorithm to triangle counting based on a \emph{filtering-and-joining} strategy, achieving a speedup of 9--260$\times$ over a sequential CPU implementation.
        \item We develop a best-of-class intersection operator and integrate it into our proposed set-intersection-based method that incorporates several optimizations yielding a speedup as high as 630$\times$ over the sequential CPU implementation and 5.1$\times$ over the previous state-of-the-art GPU exact triangle counting implementation.
        \item We formulate a parallel matrix-multiplication-based method on the GPU that achieves 5$\times$--26$\times$ speedup over the sequential CPU implementation.
        \item We provide detailed experimental evaluations of our proposed algorithms by comparing with theoretical bounds on real-world network data as well as existing best-of-class GPU and CPU implementations.

% The rest of the paper is structured as follows.
\end{enumerate}

\section{Related Works}
\label{sec:related}
We investigate three approaches---subgraph matching, set intersection, and
matrix multiplication methods---to implement exact triangle counting.
\paragraph{Subgraph Matching Algorithms} Many state-of-the-art subgraph
matching algorithms use a backtracking strategy. The first method that
leverages this technique is proposed by Ullmann~\cite{Ullmann:1976:ASI}. His
basic idea is to incrementally compose partial solutions and discard the
completed queries. Later, VF2~\cite{Cordella:2004:SIA},
QuickSI~\cite{Shang:2008:TVH}, GraphQL~\cite{He:2008:GQL},
GADDI~\cite{Zhang:2009:GDI}, and SPath~\cite{Zhao:2010:GQO} have been proposed
as improvements to Ullmann's method. These algorithms exploit different
filtering rules, joining orders, and auxiliary information to improve
performance. Lee et al.~\cite{Lee:2012:ICS} give an in-depth comparison of the
above algorithms. For large-scale graphs, Sun et al.~\cite{Sun:2012:ESM} have
proposed a parallel method to decompose the query graphs into STwigs (two-level
trees), and perform graph exploration and join strategies in a distributed
memory cloud. More recently, Tran et al.~\cite{Tran:2015:FSM} propose
a state-of-the-art parallel method using GPUs to both take advantage of the
computation power and avoid high communication costs between distributed
machines.  GPUs with massively parallel processing architectures have been
successfully leveraged for fundamental graph operations. Traditional
backtracking approaches for subgraph matching, however, cannot efficiently be
adapted to GPUs due to two problems. First, GPU operations are based on warps
(which are groups of threads to be executed in single-instruction-multiple-data
fashion), so different execution paths generated by backtracking algorithms
may cause a warp divergence problem. Second, irregular memory access patterns cause memory accesses to become uncoalesced, which degrades GPU performance.
Our algorithm is an optimized and simplified version of the method of Tran et
al.~\cite{Tran:2015:FSM}.
\paragraph{Set Intersection Algorithms} Set intersection of two sorted arrays
on the GPU is a well-studied research
problem~\cite{Ao:2011:EPL,Green:2014:FTC}. Previous research mainly focuses on
computing only one intersection set for two large sorted arrays. Green et al.'s
GPU triangle counting algorithm~\cite{Green:2014:FTC} does set intersection
computation for each edge in the undirected graph, and uses a modified
intersection path method based on merge path~\cite{Green:2012:GMP}, which is
considered to have the highest performance for large-sorted-array set
intersection on the GPU\@. Our implementation also includes a merge-path-based
strategy when doing set intersection for large neighbor lists, and several
other optimizations.
\paragraph{Triangle Counting using Matrix Multiplication}
The triangle counting algorithm using matrix multiplication we implement in
this paper is by Azad, Bulu\c{c}, and Gilbert~\cite{Azad:2015:PTC}. It is based
on traditional graph theoretic formulations of triangle counting that compute
intersections between neighbor lists. These serial algorithms have been
parallelized using the MapReduce framework~\cite{Cohen:2009:MTI}. If matrix
multiplication is to be used to perform triangle counting, it is natural to
consider some recent GPU advances in sparse matrix multiplication.
\vfill\eject\paragraph{Matrix Multiplication Algorithms}
In recent years, some advances have been made in GPU-based SpGEMM algorithms.
On GPUs, the standard SpGEMM used for comparison is the NVIDIA cuSPARSE
library. Liu and Vinter have designed a fast SpGEMM with a upper bound on
nonzeros and fast parallel inserts~\cite{Liu:2014:AEG}. Dalton et al.\ have
developed an SpGEMM algorithm called expansion, sorting, and compression
(ESC)~\cite{Dalton:2015:OSM}. This algorithm first expands possible nonzero
elements into an intermediary matrix, then sorts by columns and rows before
finally removing entries in duplicate positions. Since devising a novel SpGEMM
is beyond the scope of this paper, we will use the standard SpGEMM algorithm
provided by cuSPARSE\@.
\paragraph{General-Purpose Computing on the GPU}
Modern GPUs are massively parallel processors that support tens of thousands of
hardware-scheduled threads running simultaneously. These threads are organized
into blocks and the hardware schedules these blocks of threads onto hardware
cores. Efficient GPU programs have enough work per kernel to keep all hardware
cores busy; ensure that all cores and threads are fully supplied with work
(load-balancing); strive to reduce thread divergence (when neighboring threads
branch in different directions); aim to access memory in large contiguous
chunks to maximize achieved memory bandwidth (coalescing); and minimize
communication between the CPU and GPU\@. Because of their high arithmetic and
memory throughput, GPUs have been adopted to accelerate large-scale graph
processing.  Gunrock~\cite{Wang:2016:GAH} is a high-performance graph
processing framework on the GPU with a data-centric abstraction centered on
operations on a vertex or edge frontier, providing a high-level APIs for
programmers to quickly develop new graph primitives with small code size and
minimal GPU programming knowledge.

\section{Parallel Triangle Counting Algorithms}
\label{sec:ptc}
\begin{figure}[ht]
        \includegraphics[height=0.2\textwidth]{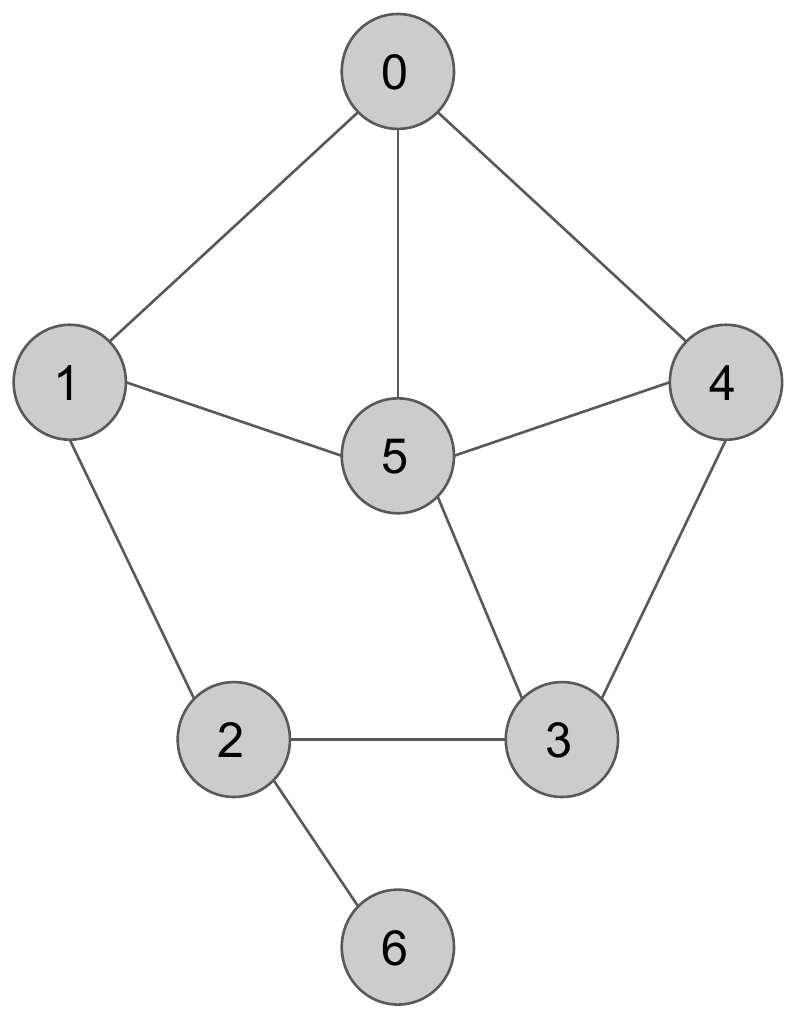}
        \centering
        \caption{A simple graph example for algorithm illustration.\label{fig:example}}
\end{figure}

In this section, we discuss the problem of triangle counting (TC) from the
perspective of subgraph matching, set intersection, and matrix multiplication.
Our implementations using these approaches all perform exact triangle counting
on undirected graphs. Figure~\ref{fig:example} will be the sample graph we use
to demonstrate our algorithms.

\subsection{TC using Subgraph Matching}
Subgraph matching is the task of finding all occurrences of a small query graph
in a large data graph. In our formulation, both graphs are undirected labeled
graphs. We can use subgraph matching to solve triangle counting problems by
first defining the query graph to be a triangle, then assigning a unified label
to both the triangle and the data graph. One advantage of using subgraph
matching to enumerate triangles in a graph is that we can get the listings of all
the triangles for free. Another advantage is that we can generalize the problem
to find the embeddings of triangles with certain label patterns.

\begin{figure}[ht]
        \includegraphics[height=0.26\textwidth]{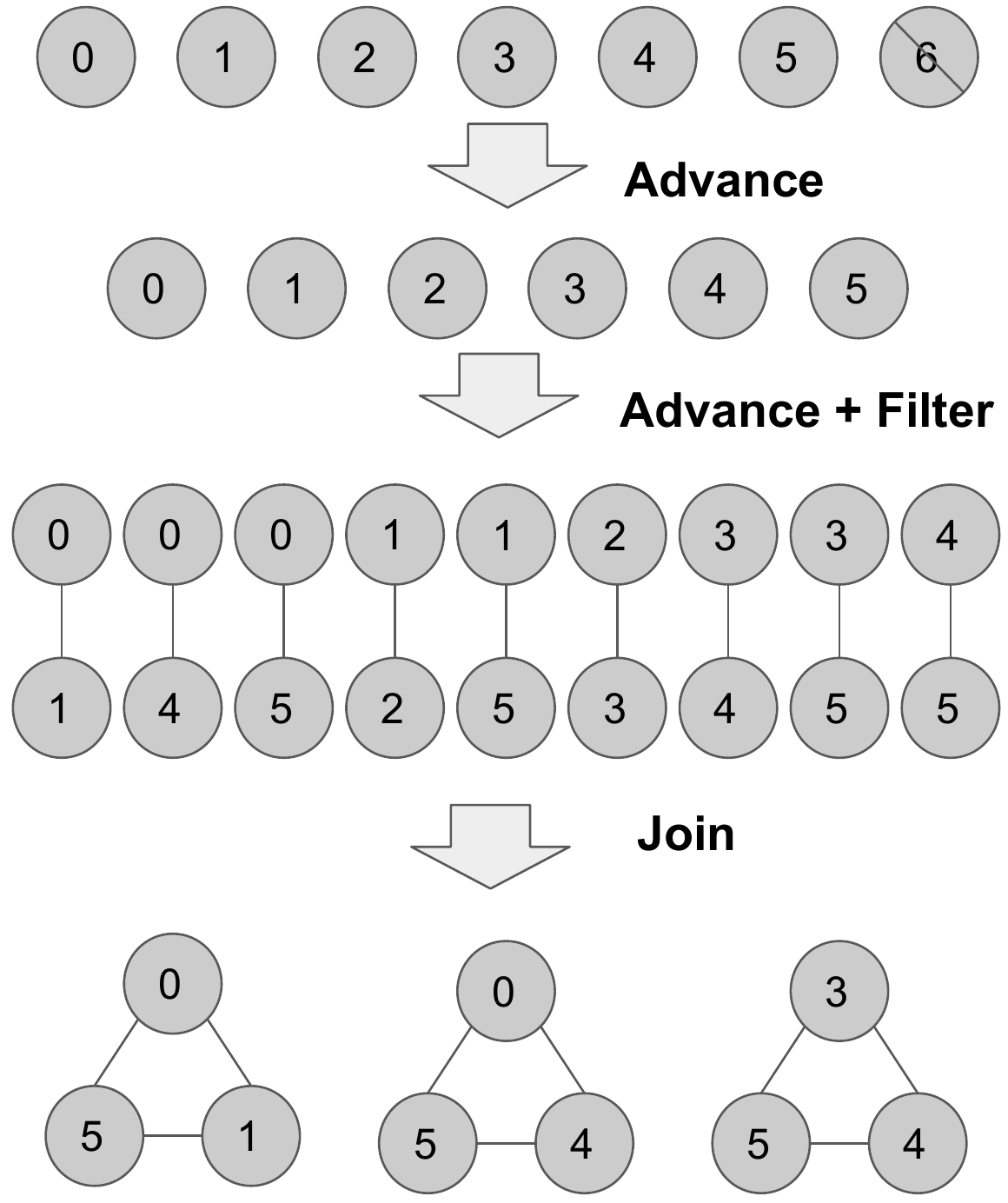}
        \centering
    \caption{A simple graph example for SM-based TC illustration.\label{fig:sm}} \end{figure}

Our method follows a filtering-and-joining procedure and is implemented using
the Gunrock~\cite{Wang:2016:GAH} programming model. Algorithm~\ref{alg:tcsm}
outlines our subgraph matching-based triangle counting implementation. In the
filtering phase, we focus on pruning away candidate vertices that cannot
contribute to the final solutions; we note that nodes with degree less than two
cannot be matched any query vertex, since every node in a triangle has a degree
of two. We use one step of Gunrock's \emph{Advance} operator to fill
a boolean candidate set (denoted as $c\_set$) for the data graph. If node $i$ in the
data graph is a candidate to node $j$ in the
query graph, then $c\_set[i][j]$ has a value of one. Otherwise, $c\_set[i][j]=0$. We then use our
$c\_set$ to label candidate edges corresponding to each query edge using another
\emph{Advance} and collect them using a \emph{Filter} operator. Then
we can get an edge list of the new data graph composed of only candidate
edges.  After collecting candidate edges for each edge in the triangle, we test
if each combination of the edges satisfies the $intersection\_rule$ defined by
the input triangle. If so, we count them into the output edge list.
Figure~\ref{fig:sm} shows a running example of the graph sample from
Figure~\ref{fig:example}.

Since every query node has the same degree of
two and the degree of a node in a large connected graph is usually larger than
two, the filtering phase cannot prune away that many nodes simply based on node
degrees and candidate neighbor lists. The most important, and also the most
time-consuming, step is joining. Tran et
al.~\cite{Tran:2015:FSM} use a two-step scheme to collect candidate edges
into a hash table. First, they count the number of candidate edges for each
query edge. Then, they compute the address of each candidate edge and assign
them to the computed address in the hash table. In our implementation, we first
use a massively parallel advance operation to label each candidate edge with
its corresponding query edge id. Then we use the \emph{select} primitive from
Merrill's \emph{CUB}
library\footnote{\label{cub}\url{http://nvlabs.github.io/cub/}}  to collect
candidate edges for each query edge.

\begin{algorithm}[!ht]\caption{TC using subgraph matching.}
\label{alg:tcsm}
\renewcommand{\algorithmicrequire}{\textbf{Input:}}
\renewcommand{\algorithmicensure}{\textbf{Output:}}
        \begin{small}
                \begin{algorithmic}[1]
                  \Require{Query Graph (Triangle) $T$, Data Graph $G$.}
           \Ensure{Number of triangles $n$ and listings of all matches.}
                        \Procedure{initialize\_candidate\_set}{$T,G$}
                        \State
                        \Call{Advance}{$G$}\Comment{Fill c\_set based on node label and degree}
                        \EndProcedure
                        \State
                        \Procedure{collect\_candidate\_edges}{$G,T,c\_set$}
                        \State
                        \Call{Advance}{$G$}\Comment{Label candidate edges with query\_edge\_id}
                        \State
                        \Call{Filter}{$G$}\Comment{Collect candidate edges}
                        \State
                        \Return{$ELIST$}
                        \EndProcedure
                        \State
                        \Procedure{Join\_candidate\_edges}{$ELIST, intersection\_rule$}
                        \State \textbf{parallel for} each candidate edge combination $\{e_i,e_j,e_k\}$

                                \If {$\{e_i,e_j,e_k\}$ satisfy $intersection\_rules$}
                                        \State Write $\{e_i,e_j,e_k\}$ to output list
                                        \State Add 1 to count
                                \EndIf
                        \State
                        \Return{count, outputlist}
                        \EndProcedure
                \end{algorithmic}
        \end{small}
\end{algorithm}

\vfill\eject\subsection{TC using Set Intersection}
The extensive survey by Schank and Wagner~\cite{Schank:2005:FCL} shows several
sequential algorithms for counting and listing triangles in undirected graphs.
Two of the best performing algorithms, \emph{edge-iterator} and \emph{forward},
both use edge-based set intersection primitives. The optimal theoretical bound
of this operation coupled with its high potential for parallel implementation
make this method a good candidate for GPU implementation.

\begin{figure}[ht]
        %\centering
        \includegraphics[width=0.5\textwidth]{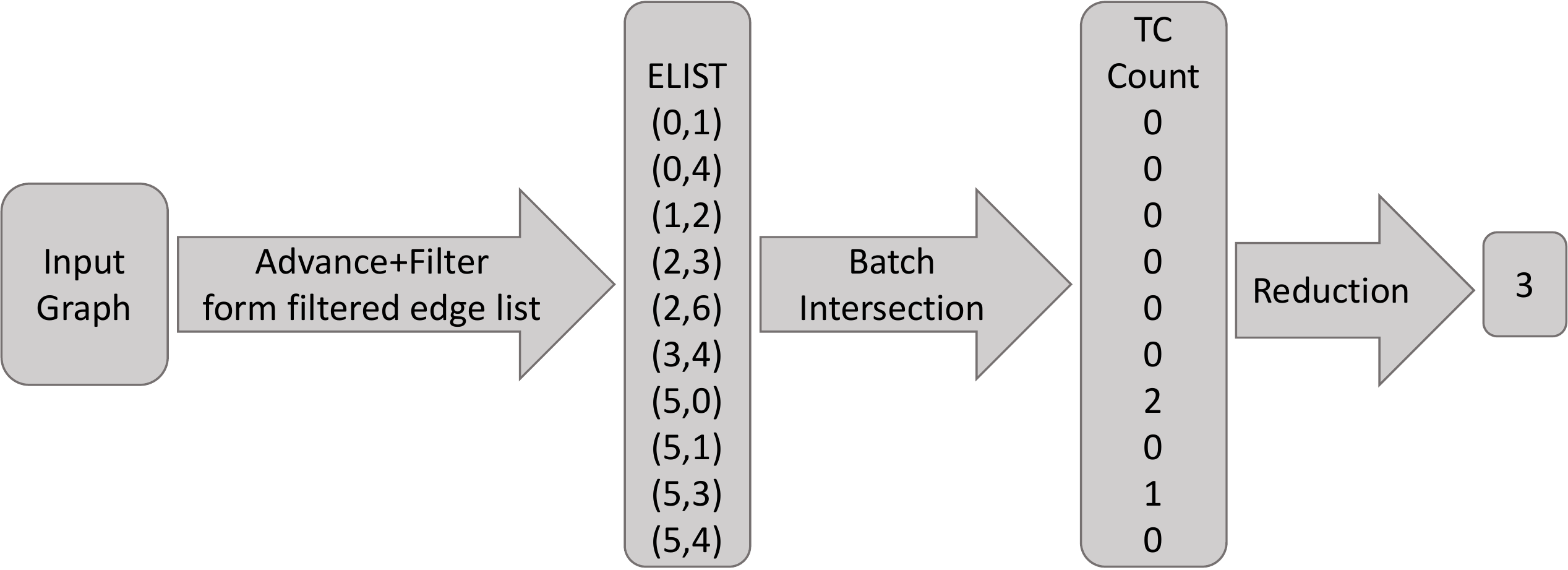}
        %\centering
        \caption{The workflow of intersection-based GPU TC algorithm.\label{fig:intersection-chart}}
\end{figure}

We can view the TC problem as a set intersection problem by the
following observation: An edge $e=(u,v)$, where $u,v$ are its two end nodes,
can form triangles with edges connected to both $u$ and $v$. Let the
intersections between the neighbor lists of $u$ and $v$ be $(w_1, w_2, \ldots,
w_N)$, where $N$ is the number of intersections.  Then the number of triangles
formed with $e$ is $N$, where the three edges of each triangle are $(u, v),
(w_i, u), (w_i, v), i \in [1,N]$. In practice, computing intersections for
every edge in an undirected graph is redundant. We visit all the neighbor lists
using Gunrock's \emph{Advance} operator. Then we filter out half of the edges by degree
order.  Thus, in general, set intersection-based TC algorithms have two stages:
(1)~forming edge lists; (2)~computing set intersections for two neighbor lists of
  an edge. Different optimizations can be applied to either stage. Our GPU
  implementation (shown in Algorithm~\ref{alg:tcintersection}) follows the
  \emph{forward} algorithm and uses several operators of the high-performance
  graph processing library Gunrock~\cite{Wang:2016:GAH}.
  Figure~\ref{fig:intersection-chart} is the flow chart that shows how this
  algorithm works on the running example (Figure~\ref{fig:example}).

\begin{algorithm}[!ht]\caption{TC using edge-based set intersection.} \label{alg:tcintersection}
    \begin{small}
        \begin{algorithmic}[1]
            \Procedure{Form\_Filtered\_Edge\_List}{$G$}
            \State
            \Call{Advance}{G}
            \State
            \Call{Filter}{G}
                        \State
                        \Return{ELIST}
            \EndProcedure
            \State
            \Procedure{Compute\_Intersection}{$G, ELIST$}
            \State
            \{SmallList,LargeList\} = \Call{Partition}{G, ELIST}
            \State
            \Call{LargeNeighborListIntersection}{LargeList}
            \State
            \Call{SmallNeighborListIntersection}{SmallList}
            \EndProcedure
            \State
            \Procedure{TC}{$G$}
            \State
            ELIST=\Call{Form\_Filtered\_Edge\_List}{$G$}
            \State
            IntersectList=\Call{Compute\_Intersection}{$G, ELIST$}
                        \State
                        Count=\Call{Reduce}{IntersectList}
                        \State
                        \Return{Count}
            \EndProcedure
        \end{algorithmic}
    \end{small}
\end{algorithm}

\subsection{TC using Matrix Multiplication}
\label{sec:ptc:mm}
The matrix multiplication formulation comes from Azad, Bulu\c{c}, and Gilbert's
algorithm for counting and enumerating triangles~\cite{Azad:2015:PTC}. Nodes in
the adjacency matrix $\mathbf{A}$ are arranged in some order. The lower
triangular matrix $\mathbf{L}$ represents all edges from row $i$ to column $k$
such that $k \leq i$. The upper triangular matrix $\mathbf{U}$ represents all
edges from row $k$ to column $j$ such that $k \leq j$. The dot product of row
$i$ of $\mathbf{L}$ with column $j$ of $\mathbf{U}$ is a count of all wedges
$i-k-j$ where $k \leq i$ and $k \leq j$. By performing a Hadamard product
(element-wise multiplication) with $A_{ij}$, we obtain an exact count of the
number of triangles $i-k-j$ for $k \leq i$ and $k \leq j$. Algorith~\ref{alg:tcmatrix}
shows the pseudocode of our GPU implementation and Figure~\ref{fig:mm}.
We illustrate this algorithm by giving a running example of the sample graph (Figure~\ref{fig:example}) in Figure~\ref{fig:mm}.

\begin{algorithm}[]
    \renewcommand{\algorithmicrequire}{\textbf{Input:}}
    \renewcommand{\algorithmicensure}{\textbf{Output:}}
    \begin{small}
    \begin{algorithmic}[1]
        \Require{Adjacency matrix $\mathbf{A}$.}
        \Ensure{Number of triangles $n$.}
        \Procedure{Triangle\_Count\_Matrix}{}
        %\Statex{}{\bf procedure} {\sc TriangleCountMatrix}
            \State{}Permute rows $\mathbf{A}$ so that it is ordered by
            an increasing number of nonzeros.
            \State{}Break matrix into a lower triangular piece
            $\mathbf{L}$ and an upper triangular piece $\mathbf{U}$ such that $\mathbf{A = L + U}$.
            \State{}Compute $\mathbf{B = LU}$.
            \State{}Compute $\mathbf{C = A \circ B}$ where $\circ$ is the Hadamard product.
            \State{}Compute $n = \frac{1}{2} \sum_{i} \sum_{j} A_{ij}$.
        %\Statex{}{\bf end procedure}
        \EndProcedure{}
    \end{algorithmic}
    \end{small}
    \caption{Counts how many triangles are in an undirected graph using matrix multiplication formulation. \label{alg:tcmatrix}}
\end{algorithm}

\begin{figure}
%\[
{\tiny \begin{eqnarray*}
\textbf{A} & = & \begin{pmatrix}
        %\centering
        %\caption{My caption}
        %\label{my-label}
        %\begin{tabular}{lllllll}
        0 & 1 & 0 & 0 & 1 & 1 & 0 \\
        1 & 0 & 1 & 0 & 0 & 1 & 0 \\
        0 & 1 & 0 & 1 & 0 & 0 & 1 \\
        0 & 0 & 1 & 0 & 1 & 1 & 0 \\
        1 & 0 & 0 & 1 & 0 & 1 & 0 \\
        1 & 1 & 0 & 1 & 1 & 0 & 0 \\
        0 & 0 & 1 & 0 & 0 & 0 & 0
        %\end{tabular}
%a & b \\
%c & d
\end{pmatrix} \\
%\]
%\[
\textbf{L} & = & \begin{pmatrix}
0 & & & & & & \\
1 & 0 & & & & \makebox(0,0){\text{\huge0}} & \\
0 & 1 & 0 & & & & \\
0 & 0 & 1 & 0 & & & \\
1 & 0 & 0 & 1 & 0 & & \\
1 & 1 & 0 & 1 & 1 & 0 & \\
0 & 0 & 1 & 0 & 0 & 0 & 0
\end{pmatrix} \\
%\end{figure}
%\begin{figure}
%\[
\textbf{U} & = & \begin{pmatrix}
0 & 1 & 0 & 0 & 1 & 1 & 0 \\
& 0 & 1 & 0 & 0 & 1 & 0 \\
& & 0 & 1 & 0 & 0 & 1\\
& & & 0 & 1 & 1 & 0 \\
& & & & 0 & 1 & 0 \\
& \makebox(0,0){\text{\huge0}} & & & & 0 & 0 \\
& & & & & & 0
\end{pmatrix} \\
%\end{figure}
%\begin{figure}
%\begin{eqnarray*}
\textbf{B} & = & \textbf{LU} \\
               & = & \begin{pmatrix}
               0 & 0 & 0 & 0 & 0 & 0 & 0 \\
               0 & 1 & 0 & 0 & 1 & 1 & 0 \\
               0 & 0 & 1 & 0 & 0 & 1 & 0 \\
               0 & 0 & 0 & 1 & 0 & 0 & 1 \\
               0 & 1 & 0 & 0 & 2 & 2 & 0 \\
               0 & 1 & 1 & 0 & 2 & 4 & 0 \\
               0 & 0 & 0 & 1 & 0 & 0 & 1
               \end{pmatrix} \\
%\end{eqnarray*}
%\begin{eqnarray*}
\textbf{C} & = & \textbf{A} \circ \textbf{B} \\
               & = & \begin{pmatrix}
                0 & 0 & 0 & 0 & 0 & 0 & 0 \\
                0 & 0 & 0 & 0 & 0 & 1 & 0 \\
                0 & 0 & 0 & 0 & 0 & 0 & 0 \\
                0 & 0 & 0 & 0 & 0 & 0 & 0 \\
                0 & 0 & 0 & 0 & 0 & 2 & 0 \\
                0 & 1 & 0 & 0 & 2 & 0 & 0 \\
                0 & 0 & 0 & 0 & 0 & 0 & 0
                \end{pmatrix}
\end{eqnarray*}}
\caption{The number of triangles is yielded by summing the elements of
\textbf{C}, then dividing by two. In this example, the number of triangles is
$\frac{6}{2} = 3$. \label{fig:mm}}
\end{figure}

\section{Implementations}
\label{sec:impl}
In this section, we discuss optimizations required for triangle counting
algorithms to run efficiently on the GPU\@.

\subsection[The Implementation of Subgraph-Matching-Based TC Algorithm]{The
	Implementation of Subgraph-Match-\\ing-Based TC Algorithm}
We make several
optimizations to the algorithm proposed by Tran et
al.~\cite{Tran:2015:FSM} in both filtering and joining phases.

\subsubsection{Optimizations for Candidate Node Filtering}
The initial filtering phase takes nodes as processing units and uses the
massively parallel \emph{advance} operation in Gunrock to mark nodes with
same labels and larger degrees as candidates in a candidate\_set. We then use
another \emph{advance} to label candidate edges based on the candidate\_set
and use a \emph{filter} operation in Gunrock to prune out all non-candidate edges and
reconstruct the graph. Note that by reconstructing the data graph, we also update
node degree and neighbor list information. So we can run the above two steps for a few
iterations in order to prune out more edges.

\subsubsection{Optimizations for Candidate Edge Joining}
The most time-consuming part in subgraph matching is joining.
Unlike previous
\emph{backtracking}-based algorithms, our joining operation forms partial
solutions in the verification phase to save a substantial amount of
intermediate space.
First, we collect candidate edges for each query edge from the new graph.
This step can be achieved by first labeling each candidate edge with its
corresponding query edge id, then using the
select primitive from Merrill's \emph{CUB} library to assign edges into query edge bins. This approach is both simpler and more
load-balanced compared to the two-step (\emph{computing}-\emph{and-assigning}) output
strategy used by Tran et al.~\cite{Tran:2015:FSM}, which requires heavy use of
gather and scatter operations. When joining the candidate edges, we verify the
intersection rules between candidate edges specified by the query graph.
We do the actual
joining only when the candidate edges satisfy the intersection rules in order to reduce the number of
intermediate results and consequently, the amount of required memory.

\subsection{The Implementation of Set-Intersection-Based TC Algorithm}

The reason behind the high performance of our set-intersection-based
implementation lies in the specific optimizations we propose in the
two stages of the algorithm.

\subsubsection{Optimizations for Edge List Generation Phase} Previous GPU
implementations of intersection-based triangle counting compute intersections
for all edges in the edge list. We believe they make this decision because of
a lack of efficient edge-list-generating and -filtering primitives; instead, in
our implementation, we leverage Gunrock's operations to implement these
primitives and increase our performance. To implement the \emph{forward}
algorithm, we use Gunrock's advance V2E operator to generate the edge list that
contains all edges, and then use Gunrock's filter operator to get rid of
$e(u,v)$ where $d(u) < d(v)$. For edges with the same $d(u)$ and $d(v)$, we
keep the edge if the vertex ID of $u$ is smaller than $v$. This efficiently
removes half of the workload. We then use segmented reduction in Merrill's
\emph{CUB} library to generate a smaller induced subgraph contains only the
edges that have not been filtered to further reduce two thirds of the workload.

\subsubsection{Optimizations for Batch Set Intersection}
High-performance batch set intersection requires a similar focus as
high-performance graph traversal: effective load-balancing and GPU utilization.
In our implementation, we use the same dynamic grouping strategy proposed in
Merrill's BFS work~\cite{Merrill:2012:SGG}. We divide the edge lists into two
groups: (1)~small neighbor lists; and (2)~large neighbor lists. We implement two
kernels (TwoSmall and TwoLarge) that cooperatively compute intersections. Our
TwoSmall kernel uses one thread to compute the intersection of a node pair. Our
TwoLarge kernel partitions the edge list into chunks and assigns each chunk to
a separate thread block.  Then each block uses the balanced path primitive from
the \emph{Modern GPU} library\footnote{Code is available at
	\url{http://nvlabs.github.io/moderngpu}} to cooperatively compute
intersections. By using this 2-kernel strategy and carefully choosing
a threshold value to divide the edge list into two groups, we can process
intersections with same level of workload together to gain load balancing and
higher GPU resource utilization.

\subsection[The Implementation of Matrix-Multiplication-Based TC Algorithm]{The
	Implementation of Matrix-Multiplic-\\ation-Based TC Algorithm}
\label{sec:schank-wagner-proof}
Our matrix-multiplication-based TC algorithm tests how well a highly
optimized matrix multiplication function from a standard GPU library (csrgemm
from cuSPARSE) performs compared to one implemented using a graph processing
library (subgraph matching and set intersection). The cuSPARSE  SpGEMM function
allocates one CSR row in the left input matrix for each thread, which performs
the dot product in linear time. The $n$ rows need to be multiplied with $n$
columns as mentioned in Section~\ref{sec:ptc:mm}. This leads to a $O(\sum_{v
	\in V}d(v)^{2})$ complexity, which is the same as the set-intersection-based TC
algorithm. For a detailed proof that shows this equivalence, see Schank and Wagner~\cite{Schank:2005:FCL}.

We compute the Hadamard product by assigning a thread to each row, which then
performs a multiplication by the corresponding value in adjacency matrix
$\textbf{A}$. We incorporate two optimizations: (1)~compute only the upper
triangular values, because the output of the Hadamard product is known to be
symmetric; (2)~use a counter to keep track of the number of triangles found by
each thread, allowing the sum of the number of triangles in each column with a single
parallel reduction.
Step 2 combines lines 5 and 6 in Algorithm~\ref{alg:tcmatrix} so that
the reduction can be performed without writing the nonzeros into global memory.

\section{Experiments and Discussion}
\label{sec:exp}
We compare the performance of our algorithms to three different exact triangle
counting methods: Green et al.'s state-of-the-art GPU
implementation~\cite{Green:2014:FTC} that runs on an NVIDIA K40c GPU\@, Green
et al.'s multicore CPU implementation~\cite{Green:2014:LBC}, and Shun et al.'s
multicore CPU implementation~\cite{Shun:2015:MTC}. Both of the state-of-the-art CPU
implementations are tested on a 40-core shared memory system with two-way
hyper-threading; their results are from their publications~\cite{Green:2014:LBC,Shun:2015:MTC}. Currently, our algorithms target NVIDIA GPUs only.
However, we hope the discoveries made in this paper can help with triangle
counting implementations on other highly parallel processors. Our CPU baseline
is an implementation based on the \emph{forward} algorithm by Schank and
Wagner~\cite{Schank:2005:FCL}.

\emph{Datasets.} We test our implementations using a variety of
real-world graph datasets from the DIMACS10 Graph Challenge~\cite{DIMACS10} and the Stanford Network Analysis Project (SNAP)~\cite{snapnets}.
Table~\ref{tab:dataset} describes the datasets. The topology of
these datasets spans from regular to scale-free.

\emph{Environment.} We ran experiments of our three implementations in this paper on a Linux workstation with
2$\times$3.50~GHz Intel 4-core, hyperthreaded E5-2637 v2 Xeon CPUs, 528~GB of
main memory, and an NVIDIA K40c GPU with 12~GB on-board memory.  Our GPU programs were compiled with NVIDIA's nvcc compiler (version~7.5.17) with the -O3 flag.

\begin{table}
  \small
  \centering
  \setlength{\tabcolsep}{3pt}
  \begin{tabular}{*{5}{c}} \toprule Dataset &Vertices&Edges&Max Degree &Type\\
    \midrule
       coAuthorsCiteseer & 227,320 & 1,628,268 & 1372 & rs
    \\ coPapersDBLP & 540,486 & 30,491,458 & 3299 & rs
    \\ road\_central & 14,081,816 & 33,866,826 & 8 & rm
    \\ soc-LJ & 4,847,571 & 137,987,546 & 20,292 & rs
    \\ cit-Patents & 3,774,768 & 33,037,896 & 770 & rs
    \\ com-Orkut & 3,072,441 & 234,370,166 & 33,007 & rs
    \\ \bottomrule
  \end{tabular}
  \caption{Dataset Description Table. The edge number shown is the number of
      directed edges when the graphs are treated as undirected graphs (if two nodes
      are connected, then there are two directed edges between them) and
      de-duplicate the redundant edges. Graph types are: r: real-world, s:
  scale-free, and m: mesh-like.\label{tab:dataset}}
\end{table}

\begin{figure*}[ht]
    \centering
    \begin{minipage}{\textwidth}
    \centering
    \includegraphics[width=0.985\textwidth]{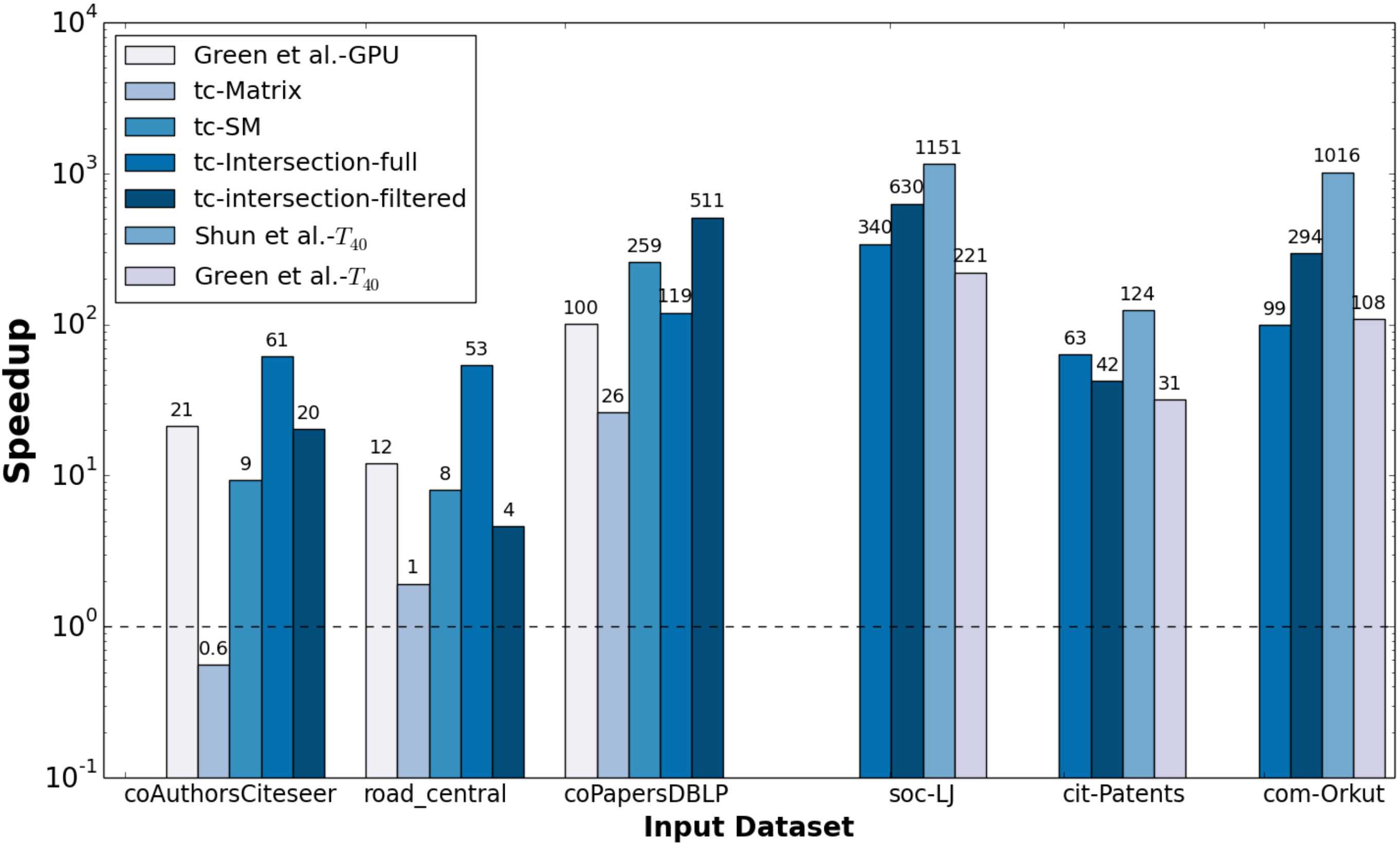}
    \end{minipage}
    \centering
     \begin{tabular}{*{7}{c}} \toprule Dataset &coAuthorsCiteseer&road\_central & coPapersDBLP & soc-LJ & cit-Patents & com-Orkut\\
    \midrule
      Baseline time (s) &0.275& 3.254 & 100.654 &564.267 & 9.856 & 1950.837
    \\ \bottomrule
  \end{tabular}
  \caption{ Execution-time speedup (top figure) for our four GPU
    implementations (``tc-Matrix'', ``tc-SM'',
    ``tc-intersection-full'', and ``tc-intersection-filtered''), Green
    et al.'s GPU implementation~\cite{Green:2014:FTC} (``Green et
    al.-GPU''), Shun et al.'s 40-core CPU
    implementation~\cite{Shun:2015:MTC} (``Shun et al.-$T_{40}$'') and
    Green et al.'s 40-core CPU implementation~\cite{Green:2014:LBC}
    (``Green et al.-$T_{40}$''). All are normalized to a baseline CPU
    implementation~\cite{Schank:2005:FCL} on six different datasets.
    Baseline runtime (in seconds) given in the
    table.\label{fig:speedup}}
\end{figure*}

\begin{figure*}[ht]
  \centering
  \begin{minipage}[l]{1.0\columnwidth}
  \centering
  \includegraphics[width=\columnwidth]{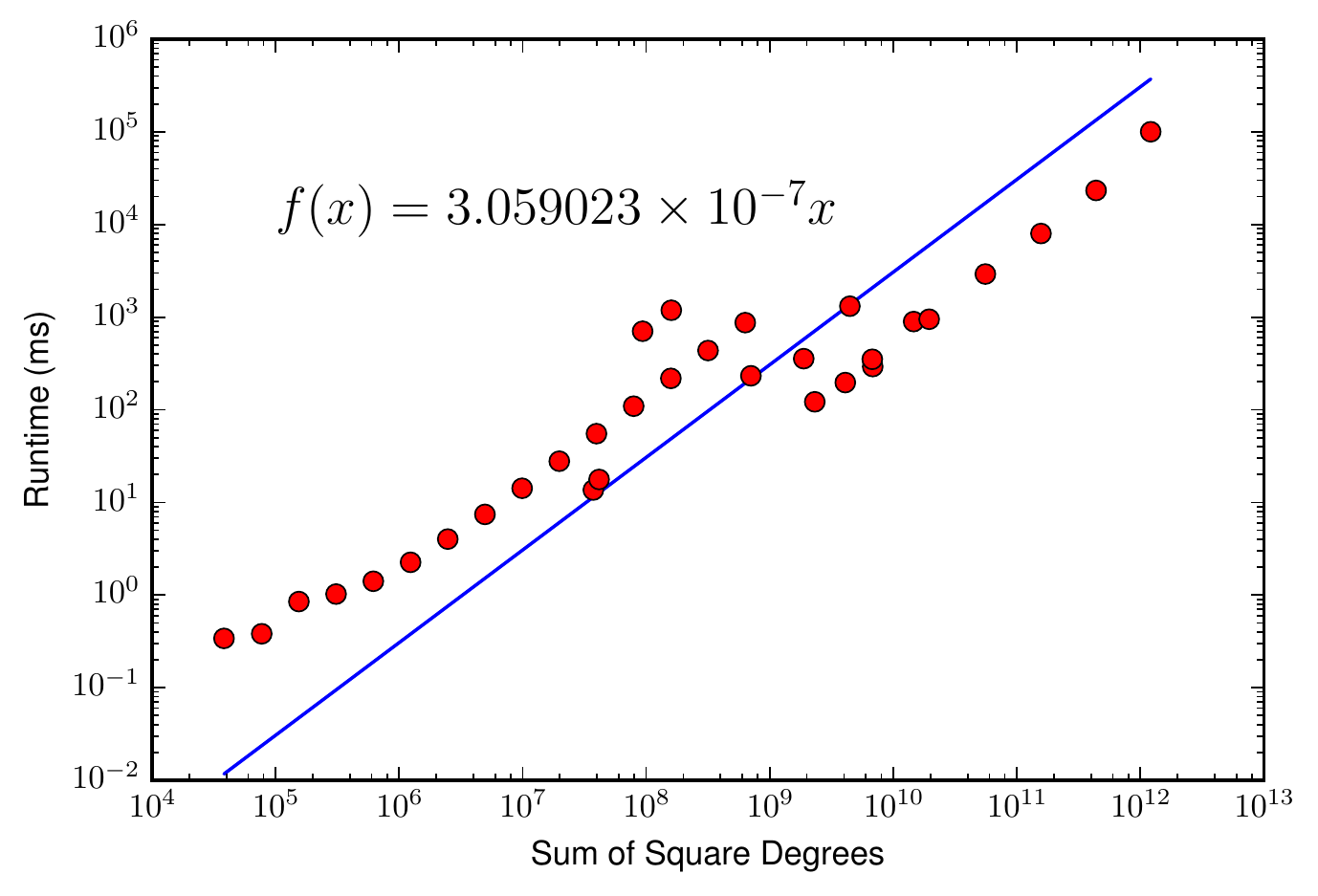}
%  \caption{Throughput on plain text (``enwik8'') as dataset size scales.\label{fig:enwik}}
  \end{minipage}
  \begin{minipage}[l]{1.0\columnwidth}
  \centering
  \includegraphics[width=\columnwidth]{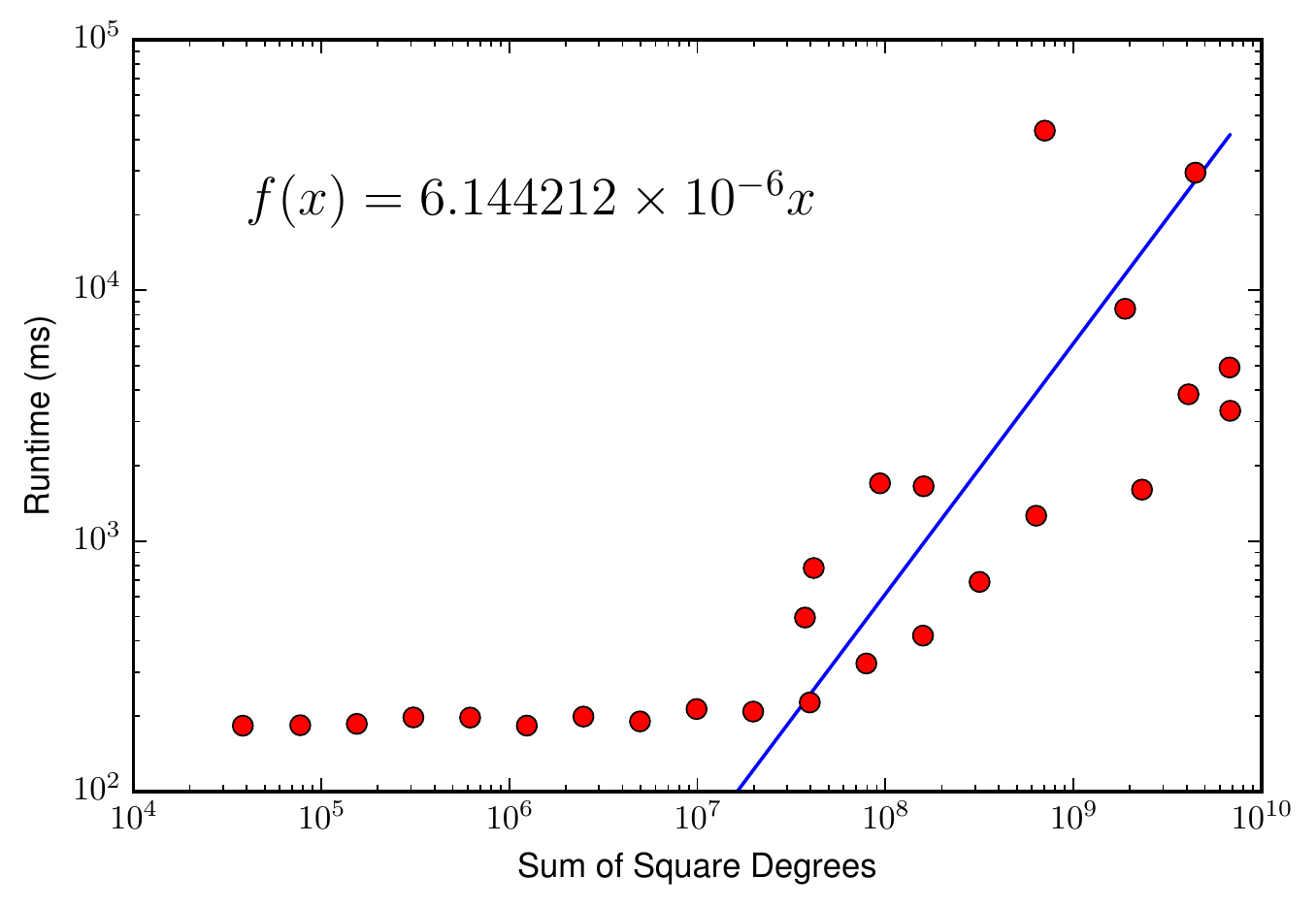}
%  \caption{Throughput on a dataset consisting only of the repeated
%    letter `A'\@. At a dataset size of 10~MB, libdivsufsort achieves
%    250~MB/s.\label{fig:special}}
  \end{minipage}
  \caption{Schank and Wagner~\cite{Schank:2005:FCL} show that the
    runtime of triangle counting on a particular graph should be proportional to the sum of
    square degrees (SSD) of that graph: $O(\sum_{v \in V}d(v)^{2})$.
    For our intersection-based (left) and matrix-based (right)
    implementations, this is true; the line of best fit for both
    runtime vs.\ SSD plots has a slope of one. The flat runtime
    baseline in the right figure indicates kernel launch overhead in
    the matrix multiplication operation. The  $y$-intercept in a
    log-log plot is equivalent to the constant term in Big-O notation.
    The leading constant term ($3.059 \times 10^{-7}$ vs. $6.144
    \times 10^{-6}$) indicates our implementation of
    intersection-based triangle counting is faster than our
    matrix-based implementation.\label{fig:ssd}}
\end{figure*}

\paragraph{Perf.\ Overview}
Figure~\ref{fig:speedup} compares our three TC implementations and several
state-of-the-art shared-memory and GPU TC implementations with a baseline
CPU implementation. In general, our intersection-based method shows better
performance than the state-of-the-art GPU implementations because of our
key optimizations on workload reduction and GPU resource utilization.
It achieves comparable performance to the fastest shared-memory CPU TC
implementation. Our matrix-based method illustrates some inherent problems with using matrix multiplication to compute TC\@. Our
subgraph-matching-based method shows some
performance wins on mesh-like datasets since we prune out a lot of leaf nodes
in our filtering base and can thus can reduce around one-third of work for mesh-like graphs
during the joining phase. Figure~\ref{fig:ssd} confirms our
expectation from Section~\ref{sec:schank-wagner-proof} that both our
intersection-based and matrix-based implementations have runtime proportional to the sum of square degrees of their input graphs.

\paragraph{Perf.\ Summary for Subgraph Matching-based TC}
Our subgraph matching-based triangle counting enumerates all the triangles for free
and thus needs a lot of memory in the joining phase. The optimizations of the joining phase
give us a performance win compared with previous state-of-the-art
subgraph matching algorithms.  As shown in Figure~\ref{fig:speedup},
most of the time, subgraph matching is slower than intersection-based
methods. Subgraph matching, however does well on some mesh-like datasets that contain a comparatively
large number of leaf nodes, which we can filter out in our first stage. For the joining phase,
the more candidate edges we get, the more edge combinations we see.
When the number of combinations exceeds the number of parallel
processors we have, each processor needs to deal with multiple
combinations sequentially, which hurts performance.

\paragraph{Perf.\ Summary for Intersection-based TC} Our
intersection-based TC with only the simple per-thread batch set intersection
kernel achieves a 2.51$\times$ speedup (geometric-mean) speedup compared to Green et
al.'s CPU implementation~\cite{Green:2014:LBC} and a 4.03$\times$
geomean speedup compared to Green et al.'s GPU
implementation~\cite{Green:2014:FTC}. We believe our speedup is the result of
two aspects of our implementation: using filtering in edge list generation and
reforming the induced subgraph with only the edges not filtered which reduces
five-sixths of the workload, and dynamic grouping helps maintain a high GPU
resource utilization.  In practice we observe that for scale-free graphs, the
last step of optimization to reform the induced subgraph show a constant
speedup over our intersection method without this step. However, for road
networks and some small scale-free graphs, the overhead of using segmented
reduction will cause a performance drop.  We expect further performance gains
from future tuning of launch settings for the large-neighbor-list intersection
kernel, which we do not believe is quite optimal yet. Furthermore, we believe that
a third kernel to do batch set intersection between one large and one small
neighbor list will continue to improve performance. This third kernel would scan each node in the smaller list and search the larger list. Highly skewed scale-free
graphs where most intersections are between a very small neighbor list and the
same large neighbor list which covers more than half of the nodes in the graph would see a performance improvement from a kernel like this. With little modification, we expect our method could also be used to compute
clustering coefficient and transitivity metrics.

\paragraph{Perf.\ Summary for Matrix Multiplication-based TC}
Figure~\ref{fig:speedup} shows how the matrix multiplication-based
TC compares with other algorithms. It has similar performance to the reference CPU algorithm in two out ot three datasets. For the largest dataset of the three  ``coPapersDBLP'', it does perform $26\times$ better, indicating better scaling.

Figure~\ref{fig:ssd} shows how matrix-based TC scales. The linear relationship beyond the high slope near
the start indicates that the algorithm is bounded by the same complexity as the
intersection-based algorithm. The low slope at the start is indicative of the
large overhead from the SpGEMM algorithm, which is mitigated once the graph
grows to a certain size.

Testing with profiler tools indicate SpGEMM is the bottleneck of our
implementation. Since sparse matrix multiplication needs to generate an
intermediate array, our results suggest that even an optimized, out-of-the-box
SpGEMM cannot compete with a graph processing library, because SpGEMM is doing
unnecessary work and within the matrix formulation. There do not seem to
be express some TC-specific optimizations. For the latter, these include: (1)~only
calculate the upper triangular of the matrix multiplication; (2)~avoid
multiplications where $\mathbf{A}$ is known to be zero \emph{a priori}; and (3)~avoid writing the output of the matrix multiplication into global memory.
Techniques such as masking the input adjacency matrix $\mathbf{A}$ have been
shown to yield speed-ups on distributed systems~\cite{Azad:2015:PTC}.
Implementing this on a GPU is a good direction for future research.

\section{Conclusion}
\label{sec:conc}
Previous work in triangle counting on massively parallel processors
like GPUs has concentrated on single implementations using a
particular methodology. With this work we analyzed three different
GPU implementations based on subgraph matching, set intersection, and
matrix multiplication. Our intersection-based implementation achieves state-of
-the-art performance and shows potential to gain even better performance.
Our matrix-based implementation shows that SpGEMM is the performance bottleneck
on GPU due to its unavoidable redundant work, although future work on
making customized modifications to reduce redundant work could lead to
better speedups. Our subgraph matching method, despite its memory bound, shows
good performance on mesh like graphs that contain a large part of leaf nodes.
Its generality also allows programmers to easily extend this method to match
more complicated subgraph patterns.

We also note that some optimizations such as filtering edge lists and improving
set intersection can easily address other problems on the GPU,
while others such as joining order selection and reducing intermediate results 
are considerably more
challenging. We hope that the most challenging ones will serve as interesting
future work for the authors and, more importantly, interesting future
work for the communities developing frameworks for those particular methodologies.

\section*{Acknowledgements}
\label{sec:acks}
Thanks to Seshadhri Comandur for providing the CPU baseline code for comparisons.
We appreciate the funding support of
DARPA XDATA under grants US Army award W911QX-12-C-0059, DARPA STTR awards D14PC00023 and
D15PC00010 as well as the National Science Foundation under grants CCF-1017399 and OCI-1032859, and a Lawrence Berkeley
Laboratory internship. Thanks also to NVIDIA for equipment
donations and server time.

\bibliographystyle{abbrv}
\bibliography{tc}

\begin{thebibliography}{10}

\bibitem{Ao:2011:EPL}
N.~Ao, F.~Zhang, D.~Wu, D.~S. Stones, G.~Wang, X.~Liu, J.~Liu, and S.~Lin.
\newblock Efficient parallel lists intersection and index compression
  algorithms using graphics processing units.
\newblock {\em Proceedings of the VLDB Endowment}, 4(8):470--481, May 2011.

\bibitem{Azad:2015:PTC}
A.~Azad, A.~Bulu\c{c}, and J.~Gilbert.
\newblock Parallel triangle counting and enumeration using matrix algebra.
\newblock In {\em IEEE International Parallel and Distributed Processing
  Symposium Workshop}, IPDPSW 2015, pages 804--811, 2015.

\bibitem{DIMACS10}
{Center for Discrete Mathematics \& Theoretical Computer Science}.
\newblock 10th {DIMACS} implementation challenge---graph partitioning and graph
  clustering.
\newblock \url{http://www.cc.gatech.edu/dimacs10/index.shtml}, Feb. 2011.

\bibitem{Cohen:2009:MTI}
J.~Cohen.
\newblock Graph twiddling in a {MapReduce} world.
\newblock {\em Computing in Science \& Engineering}, 11(4):29--41, 2009.

\bibitem{Coppersmith:1987:MMV}
D.~Coppersmith and S.~Winograd.
\newblock Matrix multiplication via arithmetic progressions.
\newblock In {\em Proceedings of the Nineteenth Annual ACM Symposium on Theory
  of Computing}, STOC '87, pages 1--6, 1987.

\bibitem{Cordella:2004:SIA}
L.~P. Cordella, P.~Foggia, C.~Sansone, and M.~Vento.
\newblock A (sub)graph isomorphism algorithm for matching large graphs.
\newblock {\em IEEE Transactions on Pattern Analysis and Machine Intelligence},
  26(10):1367--1372, Oct. 2004.

\bibitem{Dalton:2015:OSM}
S.~Dalton, L.~Olson, and N.~Bell.
\newblock Optimizing sparse matrix-matrix multiplication for the {GPU}.
\newblock {\em ACM Transactions on Mathematical Software (TOMS)}, 41(4):25,
  2015.

\bibitem{Talya:2015:ACT}
T.~Eden, A.~Levi, and D.~Ron.
\newblock Approximately counting triangles in sublinear time.
\newblock {\em CoRR}, abs/1504.00954, 2015.

\bibitem{Green:2012:GMP}
O.~Green, R.~McColl, and D.~A. Bader.
\newblock {GPU} merge path: A {GPU} merging algorithm.
\newblock In {\em Proceedings of the 26th ACM International Conference on
  Supercomputing}, ICS '12, pages 331--340, 2012.

\bibitem{Green:2014:LBC}
O.~Green, L.-M. Mungu\'{\i}a, and D.~A. Bader.
\newblock Load balanced clustering coefficients.
\newblock In {\em Proceedings of the First Workshop on Parallel Programming for
  Analytics Applications}, PPAA '14, pages 3--10, 2014.

\bibitem{Green:2014:FTC}
O.~Green, P.~Yalamanchili, and L.-M. Mungu\'{\i}a.
\newblock Fast triangle counting on the {GPU}.
\newblock In {\em Proceedings of the Fourth Workshop on Irregular Applications:
  Architectures and Algorithms}, IA3 '14, pages 1--8, 2014.

\bibitem{He:2008:GQL}
H.~He and A.~K. Singh.
\newblock Graphs-at-a-time: Query language and access methods for graph
  databases.
\newblock In {\em Proceedings of the 2008 ACM SIGMOD International Conference
  on Management of Data}, SIGMOD '08, pages 405--418, 2008.

\bibitem{Madhav:2012:FBP}
M.~Jha, C.~Seshadhri, and A.~Pinar.
\newblock From the birthday paradox to a practical sublinear space streaming
  algorithm for triangle counting.
\newblock {\em CoRR}, abs/1212.2264, 2012.

\bibitem{Kolda:2013:CTM}
T.~G. Kolda, A.~Pinar, T.~Plantenga, C.~Seshadhri, and C.~Task.
\newblock Counting triangles in massive graphs with mapreduce.
\newblock {\em CoRR}, abs/1301.5887, 2013.

\bibitem{Lee:2012:ICS}
J.~Lee, W.-S. Han, R.~Kasperovics, and J.-H. Lee.
\newblock An in-depth comparison of subgraph isomorphism algorithms in graph
  databases.
\newblock In {\em Proceedings of the 39th International Conference on Very
  Large Data Bases}, PVLDB'13, pages 133--144. VLDB Endowment, 2013.

\bibitem{snapnets}
J.~Leskovec and A.~Krevl.
\newblock {SNAP Datasets}: {Stanford} large network dataset collection.
\newblock \url{http://snap.stanford.edu/data}, June 2014.

\bibitem{Liu:2014:AEG}
W.~Liu and B.~Vinter.
\newblock An efficient {GPU} general sparse matrix-matrix multiplication for
  irregular data.
\newblock In {\em Proceedings of the 28th International Parallel and
  Distributed Processing Symposium}, IPDPS 2014, pages 370--381, May 2014.

\bibitem{Polak:2015:CTL}
A.~Polak.
\newblock Counting triangles in large graphs on {GPU}.
\newblock {\em CoRR}, abs/1503.00576, 2015.

\bibitem{Schank:2005:FCL}
T.~Schank and D.~Wagner.
\newblock Finding, counting and listing all triangles in large graphs, an
  experimental study.
\newblock In {\em Proceedings of the 4th International Conference on
  Experimental and Efficient Algorithms}, WEA'05, pages 606--609, 2005.

\bibitem{Shang:2008:TVH}
H.~Shang, Y.~Zhang, X.~Lin, and J.~X. Yu.
\newblock Taming verification hardness: An efficient algorithm for testing
  subgraph isomorphism.
\newblock {\em Proceedings of the VLDB Endowment}, 1(1):364--375, Aug. 2008.

\bibitem{Shun:2015:MTC}
J.~Shun and K.~Tangwongsan.
\newblock Multicore triangle computations without tuning.
\newblock In {\em IEEE 31st International Conference on Data Engineering},
  pages 149--160, April 2015.

\bibitem{Sun:2012:ESM}
Z.~Sun, H.~Wang, H.~Wang, B.~Shao, and J.~Li.
\newblock Efficient subgraph matching on billion node graphs.
\newblock {\em Proceedings of the VLDB Endowment}, 5(9):788--799, May 2012.

\bibitem{Tran:2015:FSM}
H.-N. Tran, J.-j. Kim, and B.~He.
\newblock Fast subgraph matching on large graphs using graphics processors.
\newblock In M.~Renz, C.~Shahabi, X.~Zhou, and A.~M. Cheema, editors, {\em
  Proceedings of the 20th International Conference on Database Systems for
  Advanced Applications}, DASFAA 2015, pages 299--315. Springer International
  Publishing, Cham, Apr. 2015.

\bibitem{Tsourakakis:2009:DCT}
C.~E. Tsourakakis, U.~Kang, G.~L. Miller, and C.~Faloutsos.
\newblock {DOULION}: Counting triangles in massive graphs with a coin.
\newblock In {\em Proceedings of the 15th ACM SIGKDD International Conference
  on Knowledge Discovery and Data Mining}, KDD '09, pages 837--846, 2009.

\bibitem{Ullmann:1976:ASI}
J.~R. Ullmann.
\newblock An algorithm for subgraph isomorphism.
\newblock {\em Journal of the ACM}, 23(1):31--42, Jan. 1976.

\bibitem{Wang:2012:TDM}
J.~Wang and J.~Cheng.
\newblock Truss decomposition in massive networks.
\newblock {\em Proceedings of the VLDB Endowment}, 5(9):812--823, May 2012.

\bibitem{Wang:2010:OTB}
N.~Wang, J.~Zhang, K.-L. Tan, and A.~K. Tung.
\newblock On triangulation-based dense neighborhood graph discovery.
\newblock {\em Proceedings of the VLDB Endowment}, 4(2):58--68, 2010.

\bibitem{Wang:2016:GAH}
Y.~Wang, A.~Davidson, Y.~Pan, Y.~Wu, A.~Riffel, and J.~D. Owens.
\newblock {G}unrock: A high-performance graph processing library on the {GPU}.
\newblock In {\em Proceedings of the 21st ACM SIGPLAN Symposium on Principles
  and Practice of Parallel Programming}, PPoPP 2016, pages 11:1--11:12, Mar.
  2016.

\bibitem{Watts:1998:CDS}
D.~J. Watts and S.~H. Strogatz.
\newblock Collective dynamics of `small-world' networks.
\newblock {\em Nature}, 393(6684):440--442, 1998.

\bibitem{Zhang:2009:GDI}
S.~Zhang, S.~Li, and J.~Yang.
\newblock {GADDI}: Distance index based subgraph matching in biological
  networks.
\newblock In {\em Proceedings of the 12th International Conference on Extending
  Database Technology: Advances in Database Technology}, EDBT '09, pages
  192--203, 2009.

\bibitem{Zhao:2010:GQO}
P.~Zhao and J.~Han.
\newblock On graph query optimization in large networks.
\newblock {\em Proceedings of the VLDB Endowment}, 3(1--2):340--351, Sept.
  2010.

\end{thebibliography}

\end{document}